\documentclass{article}
\usepackage{graphicx}\def\apjl{ApJL }

\def\apjs{ApJS }

\def\aap{A\&A }

\begin{document}
\begin{center} {\large\bf  Mass segregation in diverse environments \linebreak
\linebreak
Priya Hasan\thanks{E-mail:
priyashasan@yahoo.com, priya.hasan@gmail.com} 
and S N Hasan\thanks{E-mail:najam\_hasan@yahoo.com}\\
Department of Astronomy, Osmania University, Hyderabad~-~500007,~India.\\}
\end{center}

\begin{abstract}
In this paper, using 2MASS photometry, we study the mass functions $\phi(M) = dN/dM \propto M^{-\alpha}$ of a sample of nine clusters of ages varying from 4~Myr--1.2~Gyr and Galactocentric distances from 6--12~kpc. We look for evidence of mass segregation in these clusters by tracing the variation in the value of $\alpha$ in different regions of the cluster as a function of the parameter $\tau = t_{age}/t_{relax}$ (where  $t_{age}$ is the age of the cluster and $t_{relax}$ is the relaxation time of the cluster), Galactocentric distance, age and size of the cluster.  The value of $\alpha$ value increases with age and $\tau$ and fits straight lines with slopes $m$  and y-intercepts $c$ given by $m=0.40\pm0.03$, $c=-1.86\pm0.27$ and $m=0.01\pm0.001$, $c=-0.85\pm0.02$, respectively and is a clear indicator of the dynamical processes involved. The confidence level of the  Pearson's product-moment correlation of $\alpha$  with age is 0.76 with p=0.002 and with $\tau$ is 0.71 with p=0.007. The value of $\alpha$ also increases with Galactocentric distance, indicating the presence of a larger relative number of low mass stars in clusters at larger  Galactocentric distances. We find two clusters, viz. IC~1805 and NGC~1893, with evidence of primordial or early dynamical  mass segregation. Implications of primordial mass segregation on the formation of massive stars and recent results supporting early dynamical mass segregation are discussed.  
\end{abstract}

Keywords\\
star clusters: young -- near-infrared photometry -- colour--magnitude diagrams -- pre-mainsequence stars -- initial mass function--relaxation time-- 2MASS

\section{Introduction}
The distribution of mass amongst the stars born from a parent cloud is described by the initial mass function (IMF). It is a fundamental parameter not only in understanding the basic star formation process, but also in determining the properties and evolution of stellar systems, which are the basic building blocks of galaxies. The IMF estimated for different populations in which the stars can be observed individually show an extraordinary uniformity (Bastian et al. 2010). This uniformity appears to be present for stellar populations including present-day star formation in small molecular clouds, rich and dense massive star-clusters forming in giant clouds and also with old and metal-poor stellar populations that may be dominated by dark matter. The universality, origin and dependence on physical conditions of the IMF is a very active research area and is very crucial to understanding the basic physics of star formation  (Kroupa 2002; Bonnell et al. 2007). The evolution of the IMF is influenced by the evolution of individual stars, the redistribution of stars of different masses and the loss of low mass stars by evaporation. Recent studies by  (Goodwin and Kouwenhoven 2009) suggest that the same IMF can be derived from different modes of star formation and thus questioned if the IMF is a direct imprint of the star-formation process.

Star clusters are an ideal test bed for  studies of the IMF as they are a collection of coeval stars formed from the same parent cloud. Hence many uncertainties like reddening, distance, metallicity, etc in determination of stellar masses are minimised. They are suitable for  studies on star formation and the dynamics of stellar systems formation and the dynamics of stellar systems (Lynga 1982;Janes and Phelps 1994; Kharchenko et al. 2005; Friel 1995; Bonatto and Bica 2005). The term ‘ecology of star clusters’,
as coined by Heggie (1992), shows the close interplay between stellar dynamics, stellar evolution, the clusters’ stellar content and the dynamics and properties of the host galaxy all which contribute to their structure and evolution.

     Mass segregation is the distribution of stars according
to their masses, leading to the concentration of high mass
stars near the centre and the low mass ones away from the
centre. This can take place as a result of dynamical interac-
tions between stars in young clusters or could be primordial
in nature (Bonnell and Davies 1998; Gouliermis et al. 2004;
de Marchi et al. 2006; Vesperini 2010; de Grijs et al. 2002,
and references therein). For very young clusters, where the
age of these clusters is small compared to their relaxation
time, the process of dynamical segregation seems less likely,
and this timescale argument has been used as evidence that
primordial segregation has played a role (Hillenbrand and
Hartmann 1998; Bonnell and Davies 1998; Raboud and Mer-
milliod 1998). Examples of such clusters with ages less than
5 Myr include: Mon R2 (Carpenter et al. 1997); IC 1805
(Sagar et al. 1988); NGC 1893 (Sharma et al. 2007); NGC
6530 (McNamara and Sekiguchi 1986); NGC 6231 (Raboud
and Mermilliod 1998); and the Orion Nebula Cluster (ONC)
(Hillenbrand and Hartmann 1998). However, the simulations
by Moeckel and Bonnell (2009) show that for such young sys-
tems, star formation scenarios predicting primordial mass
segregation are inconsistent with observed segregation lev-
els. Recent work by Allison et al. (2009, 2010) showed that
early mass segregation can be due to dynamical eﬀects even
in timescales as short as a Myr, thus not requiring the need
of primordial mass segregation.
     Mass segregation has been studied using the variation of
the slope α of the mass function (MF) in diﬀerent regions of
clusters (Bica et al. 2006; Hasan et al. 2008). The steepness
of MF in the outer regions of the clusters compared to that
of the inner regions, indicates the presence of mass segrega-
tion in clusters. In an earlier paper, using the homogeneous
data of the Two Micron All Sky Survey (2MASS), Hasan
et al. (2008) studied a sample of four young clusters to test
if the observed mass segregation is an imprint of the star
formation process or is due to the dynamics of the clusters.
They found that the observed mass segregation of the sam-
ple of young clusters studied, could be explained on the basis
of the dynamics. It was found by Bonatto and Bica (2005);
Sharma et al. (2008), that the MF slopes (in the outer region
as well as the whole cluster) undergo an exponential decay
with the evolutionary parameter τ = tage /trelax and that
the evaporation of low-mass members from outer regions of
the clusters is not signiﬁcant at larger Galactocentric dis-
tances of 9 – 10.8 kpc. The parameter τ is an evolutionary
parameter (Bonatto and Bica 2005) which indicates the ex-
tent to which the cluster has relaxed. The relaxation time
trelax is a characteristic time during which stars in a cluster
tend to achieve equipartition of energy and the high mass
stars with lesser kinetic energy sink to the core and the low
mass stars move to the outer regions of the cluster (Binney
and Tremaine 2008).

To make inferences based on the properties and fundamental parameters of clusters, it is essential to use homogeneous samples of photometric data, coupled with uniform methods of data analysis. In this paper, we have selected a sample of nine clusters with varying ages, sizes and Galactocentric distances to study mass segregation and the change in $\alpha$ in clusters in diverse environments.  The clusters, viz. NGC~6704, NGC~6005, NGC~6200, NGC~6604, IC~1805, NGC~2286, NGC~2489, NGC~2354 and NGC 1893, are studied using photometric data from the 2MASS (Skrutskie et al. 2006). The 2MASS covers 99.99\% of the sky in the near-infrared $J$~(1.25~$\mu$m), $H$~(1.65~$\mu$m) and $K_{s}$~(2.16~$\mu$m) bands (henceforth $K_{s}$ shall be referred to as $K$). The 2MASS database has the advantages of being homogeneous, all sky (enabling the study of the outer regions of clusters where the low mass stars dominate) and covering near infrared wavelengths where young clusters can be well observed in their dusty environments. 
 Many papers devoted to the study of clusters using the 2MASS have been presented in the past few years  (Bonatto et al. 2006; Bica et al. 2003; Tadross 2008; Dutra et al. 2002)  showing the potential of this database.  We use the results of Hasan et al 2008) on four clusters  and the results of this work on nine clusters to study the dependence of $\alpha$ on $\tau$, Galactocentric distance, age and size of the cluster.  We study the structures and dynamical states of our sample of clusters and determine their MFs and degree of mass segregation in various regions of the clusters. We construct radial density profiles (RDPs), colour--magnitude diagrams (CMDs),  colour--colour diagrams (CCs), luminosity functions (LFs) and MFs.  The Galactocentric distance has been calculated based on the IAU-endorsed distance $R_{o}= 8.5$~kpc.


The plan of the paper is as follows: Section 2 describes the clusters in our sample and shows the corresponding RDPs and the values obtained for the limiting radii for these clusters. Section 3 describes the method of selecting cluster members  and the corresponding values of fundamental parameters obtained.  LFs and MFs are described in Section 4 and a comparative study of these clusters is in the concluding Section 5. 

\section{Cluster Sample}
\begin{figure}
\centering
\includegraphics[width=8cm,height=7cm]{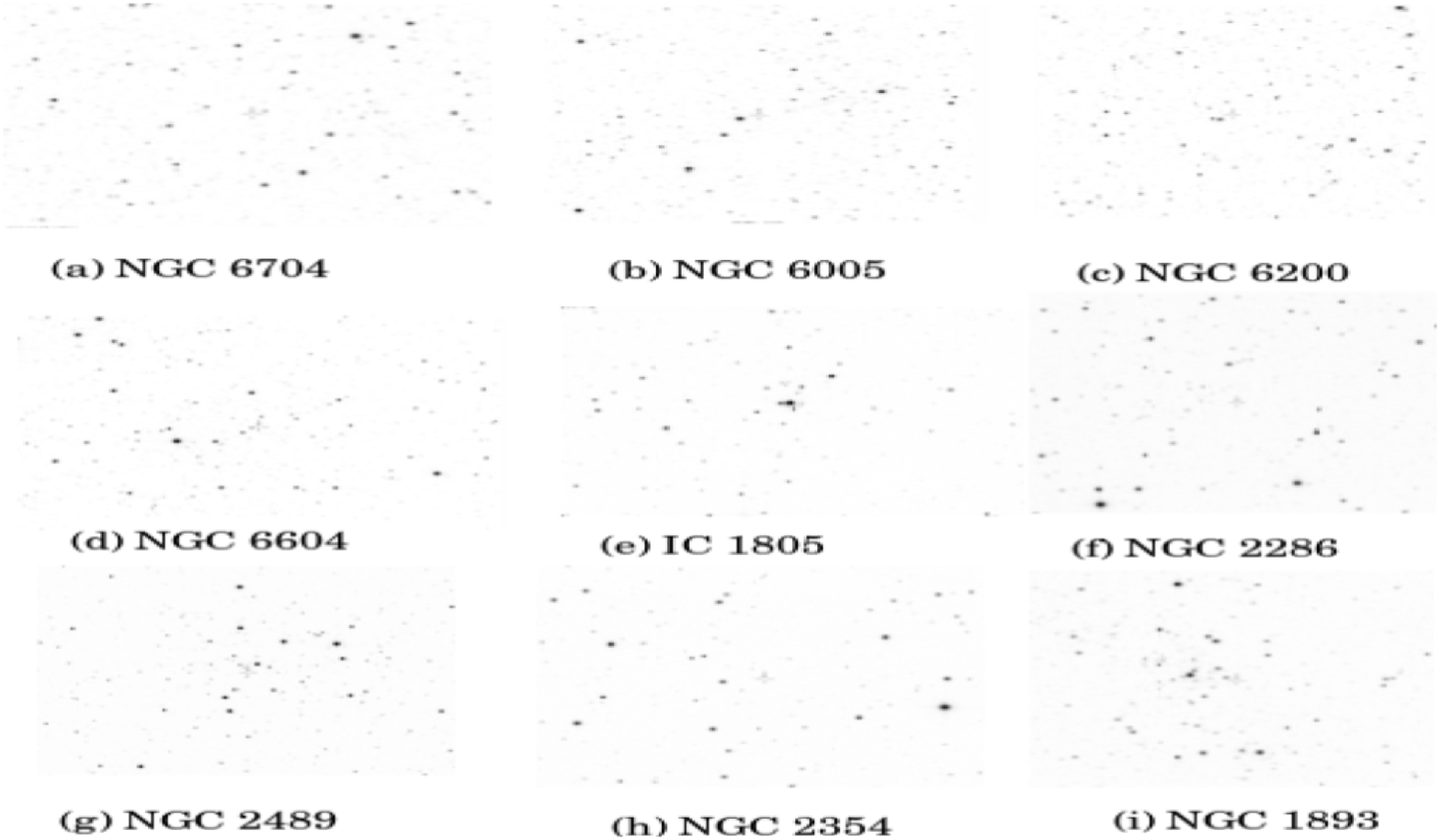}
\caption{Mosaics of 2MASS $JHK$ images of cluster areas. Field of view in arc minutes is given in brackets. In all images, North is up, East is left. (a) NGC~6704(8.84X8.84) (b) NGC~6005(11.57X8.84) (c) NGC~6200(11.57X8.84) (d) NGC~6604(11.57X8.84) (e) IC~1805(8.84X8.84) (f) NGC~2286(8.84X8.84) (g) NGC~2489(8.84X8.84) (h) NGC~2354(8.84X8.84) (i) NGC~1893(8.84X8.84).  }
\label{clusplot}
\end{figure}

 The images of the target clusters using the 2MASS  are shown in Figure \ref{clusplot}. The $JHK$ bands have been used to construct mosaics. The cluster parameters from Dias et al 2007) are given in Table~\ref{clusterdata}.  In the table, RA(2000) \& Decl.(2000) are the right ascension and declination for the epoch 2000, $l$ \& $b$ are the Galactic longitude and latitude, Ang.Dia is the angular diameter,  Distance is the distance from the Sun, $E(B-V)$ is the reddening,  log $t_{age}$ is the logarithm of the age of the cluster and $R_{GC}$ is the  Galactocentric distance. A random sample of clusters in diverse environments was selected such that it covered a range of clusters of varying age, Galactocentric distance and size.

\begin{table*}
\small

\caption{Basic cluster parameters Dias et al 2007}
\label{clusterdata}
\begin{tabular}{llllllllll}
\hline
Cluster&  RA(2000) & Decl.(2000) & $l$ &$b$ & Ang.Dia & Distance & $E(B-V)$ & log $t_{age}$ & $R_{GC}$   \\
       & h:m:s & d:m:s & deg& deg&arc min & pc & mag & log(yr) & kpc\\
           \hline
 NGC 6704 & 18 50 45 & -05 12 18 &  28.22  & -2.22  & 5  & 2974& 0.72 &7.9 & 6 \\
 NGC 6005  & 15 55 48 & -57 26 12 & 325.78  & -2.99  & 5  & 2690&  0.45  & 9.1 & 6.5\\
 NGC 6200 & 16 44 07 & -47 27 48 & 338 & -1.07 & 14 & 2054& 0.58 &6.9& 6.6\\
 NGC 6604 & 18 18 03 & -12 14 30 &  18.25  &  1.69  & 5  & 1696& 0.97 &6.8& 6.9 \\
 IC 1805  & 02 32 42 & +61 27 00 & 134.73  &  0.92  & 20 & 2344& 0.87  & 6.1 & 10.3 \\
 NGC 2286 & 06 47 40 & -03 08 54 & 215.31  & -2.27  & 14 & 2600& 0.66  & 8.3 & 10.7 \\
 NGC 2489 & 07 56 15 & -30 03 48 & 246.71  & -0.77  & 6  & 3957&  0.37 & 7.3 & 10.7\\
 NGC 2354  & 07 14 10 & -25 41 24 & 238.37  & -6.79  & 18 & 4085&  0.31 & 8.1 & 11.2 \\
 NGC 1893  & 05 22 44 &  +33 24 42& 173.59  & -1.68  & 25 & 6000&    0.45&   6.5& 14.5\\ 
 \hline
\end{tabular}
\end{table*}

NGC~6704 has been studied by Delgado et al. (1997) who found the reddening to be 0.69 with no signs of differential reddening. $BVI$ CCD photometry of NGC~6005 was presented by   Piatti et al. (1998) and the reddening was found to be $0.45\pm0.05$. NGC~6200 is a loose young open cluster in the Sagittarius-I arm extension and has been studied using $UBV$ photometry by   Fitzgerald et al. (1977) to find no obvious differential reddening. NGC~6604 has been studied by  Forbes and DuPuy (1978),   Barbon et al. (2000) and   De Becker et al. (2005). Using three independent techniques,   Barbon et al. (2000) found the mean reddening to the cluster to be $1.02\pm 0.01$ mag with no evidence for a marked differential reddening. 
IC~1805 has been studied bySagar and Yu (1990); Massey et al. (1995); Sung and Lee
(1995). Sagar and Yu (1990) found that there is a normal extinction law in the direction of the cluster.
Proper motion studies of NGC~2286 were made by  Zhao et al. (1990); Tian (1994).  The mean color excess $E(B-V)$ was found by Pan et al. (1992) to be $0.40\pm0.1$ mag. NGC~2489,  a rich open cluster in Puppis, was studied using  photographic plates by  Lindoff and Johansson (1968) and  $UBV$ measurements were made by  Ramsay and Pollaco (1992). 
    Piatti et al. (2007) found a distance of 1800~pc to this cluster with a reddening of $E(B-V)=0.30\pm0.05$ mag and age of 500~Myr. 
   $UBV$ photometry of NGC~1893 has been presented by Moffat and Vogt (1974); Massey et al. (1995). Vallenari et al.99 did near-infrared photometry of the cluster to find an age between 4-6~Myr and identified candidate pre-main sequence stars showing an infrared excess. Tapia et al. (1991) estimated the age of the cluster to be 4~Myr and derived the distance modulus 13.18$\pm$0.11 mag, and the reddening in visual magnitudes $A_{v}$ =1.68 mag. Marco et al. (2001) did $ubvyH_{\beta}$ CCD photometry of 40 very likely main-sequence (MS) members to derive reddening  $E(b-y)$ as 0.33$\pm$0.03 mag and distance modulus $V_{0}-M_{V} =13.9\pm0.2$ mag for NGC~1893. Lying in the Aur OB2 association toward the Galactic anti-centre, NGC~1893 is associated with the HII region IC~410 and is at a distance $\geq$ 11~kpc from the Galactic centre. 
A comprehensive multiwavelength study of the star-forming region NGC~1893 to explore the effects of massive stars on low-mass star formation has been made by Sharma et al (2007).

\section{Membership, Colour--Magnitude and Colour--Colour Diagrams }

VizieR was used to extract $JHK$ 2MASS photometry of the stars in a circular area of radius $30'$ from the approximate centre listed in Table~\ref{clusterdata}. We plotted the apparent CMDs for a small central area of $3'-5'$ of the cluster  (with minimum field star contamination) and used a field region of the same area to decontaminate the CMD. The point-source signal-to-noise $S/N =10$ limit for the 2MASS database is achieved at or fainter than $J=15.8$ mag, $H=15.1$ mag and $K =14.3$ mag for virtually the entire sky and hence we have used the above magnitude limits to extract the 2MASS data using Vizier\footnote{http://vizier.u-strasbg.fr/cgi-bin/VizieR?-source=II/246}. Further, we have also added the constraint that photometric errors in each band are $\leq 0.2$ mag. Completeness is also affected by source confusion or regions of high source density. The primary areas of confusion are (1) longitudes $\pm 75^0$ from the Galactic center and latitudes $\pm 1^0$ from the Galactic plane and (2) within an approximately $5^0$ radius of the Galactic center.\footnote{http://www.ipac.caltech.edu/2mass/releases/allsky/doc/explsup.html}  For clusters of our sample lying in these regions, the 99.9\% completeness limits varying with Galactic coordinates are shown in Table \ref{comp}. For all these clusters, the field star contamination is also very high and hence we do not use fainter magnitudes in our analysis.

\begin{table*}
\small
\label{comp}
\caption{Completeness Limits}
\begin{tabular}{llll}
\hline
Cluster&  J  & H  & K  \\
       & mag& mag & mag\\
           \hline
 NGC 6704 &    15.8 & 15 & 14.3 \\
 NGC 6005  &   15.8 & 14.8 &14.3 \\
 NGC 6200 &   15.8 & 14.5  &14.3\\
 NGC 6604 &    15.5  & 14.5 & 14.3\\
 \hline
\end{tabular}
\end{table*}

Clusters located towards the Galactic centre are also difficult to observe since they suffer from high interstellar absorption and/or high field star contamination and hence such clusters are a minority in catalogues. The first four clusters in our sample, i.e., NGC~6704, NGC~6005, NGC~6200 and NGC~6604 present the above difficulties and hence are  of particular interest. 

The field star decontamination procedure similar to the one applied by one applied by Bonatto et al. (2006); Bica et al. (2006); Bonatto and Bica (2007) is used to study the intrinsic cluster CMDs.  In this method, we divide the CMD into cells and count the number of stars in the field and in the cluster area. Assuming that the number of field stars is constant, we randomly remove in each cell, stars equal to the number expected in the field to obtain a `clean' cluster CMD. 
 In crowded field regions, the field star density at fainter magnitudes may be larger than that of the cluster area, thus artificially truncating the main sequence. As this method artificially removes stars and distorts the RDPs, we used this method only to uncover the cluster CMDs and colour--colour diagrams. It is used to fit the isochrones to derive the reddening and distance of the cluster. To study the cluster structure, LF and MF we use the probable members obtained by the photometric criterion Walker 1965 lying within the area of the cluster derived from the radial density profiles.  

The photometric method described by Walker (1965) involves plotting all the stars within the radius obtained using radial density profiles in the $m_{J0} Vs  M_J$ plane where $m_{J0}$ is the apparent unreddened magnitude and $M_J$ is the absolute magnitude. A  straight line representing the adopted distance modulus is drawn with boundaries of 0.75 mag which is the a maximum deviation caused by an unresolved binary with equal components. Observational scatter can cause a vertical displacement of not more than 0.5 mag for stars appearing on the main sequence.  All stars lying within these boundaries and also on the border areas are treated as members. This method is also called the evolutionary track method. A small error in estimation of the distance modulus will not lead to misidentification of a large number of members. This method identifies only main sequence stars while other luminosity classes and groups require other methods for member identification. This method has been described in detail in an earlier paper Hasan et al. 2008.

\begin{figure} 
\centering
\includegraphics[width=8cm,height=10cm]{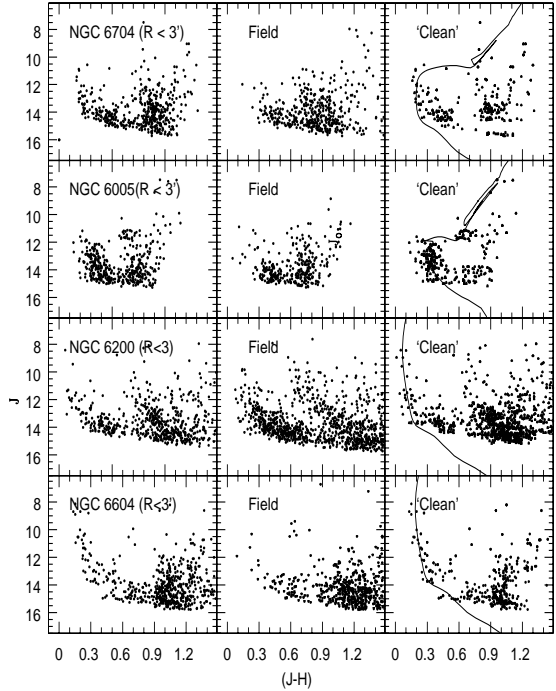}
\caption{Apparent colour--magnitude diagrams for the clusters, offset field and the `cleaned CMD' for clusters within the solar orbit: NGC~6704, NGC~6005, NGC~6200, NGC~6604. Also plotted are the isochrones Girardi et al. 2002 for the  `cleaned' CMD.}
\label{appall}
\end{figure}

\begin{figure} 
\centering
\includegraphics[width=8cm,height=10cm]{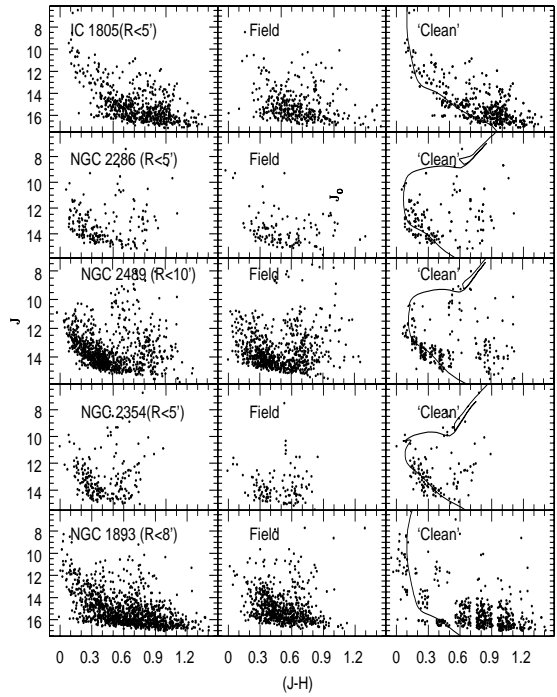}
\caption{Apparent colour--magnitude diagrams for the clusters, offset field and the `cleaned CMD'   for  clusters beyond the solar orbit: IC ~1805, NGC~2286, NGC~2489, NGC~2354 and NGC 1893. Also plotted are the isochrones Girardi et al. 2002 for the `cleaned' CMD.}
\label{appall2}
\end{figure}

The apparent CMDs for the clusters obtained by extracting stars from the central regions of the clusters,  an offset field of the same area  and the field star decontaminated or `clean' cluster CMDs are shown in the Fig. \ref{appall} and Fig \ref{appall2}. 

\begin{table*}
\small
\caption{Cluster Parameters}
\begin{tabular}{lllll}
\hline
   Cluster  & Reddening $E(B-V)$& Distance &  Age& Reference  \\
 & mag & pc & Myr & \\
             \hline
  NGC~6704 & 0.71  &1905  &20& Forbes and DuPuy (1978) \\ 
           & 0.69 & 1820& 200& Delgado et al. (1997)\\
         & 0.69 &2884 & 250& This work \\\hline
 NGC~6005& 0.45 &  2690&1200&  Piatti et al. (1998) \\
         & 0.4 & 1585& 1258&This work \\ \hline
 NGC~6200& 0.63 & 2400& -&   Fitzgerald et al. (1977)\\
         & 0.58 &2050 &6.3 &This work \\ \hline
 NGC~6604& 1.02  &1700 &5 &   Barbon et al. (2000)\\ 
         &0.97  &1700 &6.3 &This work \\ \hline
 IC~1805& 0.6 &2400 &0.25--1.5 & Sung and Lee 1995\\
         & 0.7--1.1 &1479 & 4&This work \\ \hline
 NGC~2286& 0.4 & 1510 &63 &Pan et al. (1992) \\
         &  0.3& 2618& 200&This work \\ \hline
 NGC~2489& 0.30 &1800 & 500&     Piatti et al. (2007)\\
         & 0.4 & 1445 &316 &This work \\ \hline
 NGC~2354  &0.15  & 1445 &1000 &Ahumada and Lapasset 1996\\
                   & 0.13 & 1445& 1000& Claria et al. (1999)
     \\
         &0.13  &1148 &630 &This work \\ \hline
 NGC 1893&0.4--0.6  & 3250&- & Sharma et al (2007)\\

         & 0.45--0.65 & 3630& 4&This work \\ \hline
\end{tabular}
\label{allpar}
\end{table*}

\begin{figure}
\centering
\includegraphics[width=8cm,height=7cm]{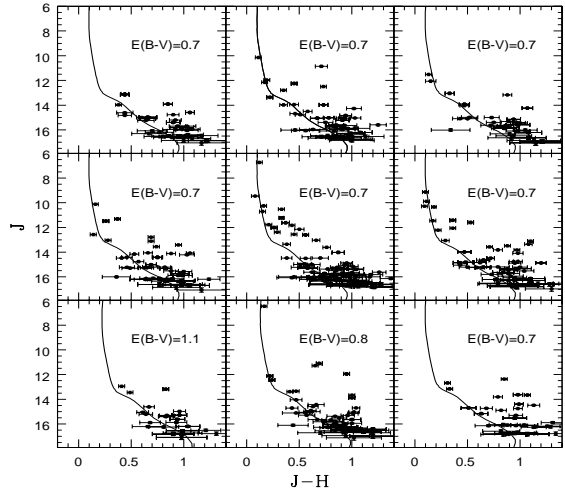}
\caption{IC 1805: Differential reddening: Values obtained by isochrone fitting for $E(B-V)$ have been indicated in the respective cells. North is up.}
\label{red1805}
\end{figure}

\begin{figure}
\centering
\includegraphics[width=8cm,height=7cm]{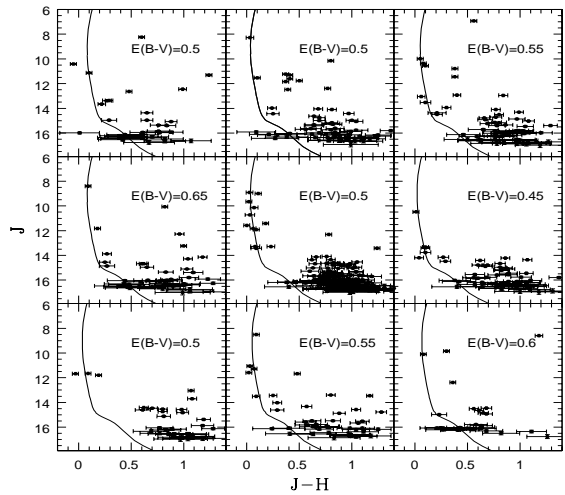}
\caption{NGC 1893: Differential reddening: Values obtained by isochrone fitting for $E(B-V)$ have been indicated in the respective cells. North is up.}. 
\label{red1893}
\end{figure}

%
%
%
%
In the case of IC~1805 and NGC~1893, which show signs of  differential reddening (Sagar and Yu 1990; Sharma et al. 2007), the entire cluster region was divided into 9 regions for which the reddening values were determined individually by isochrone fits. Stars were then corrected for their reddening values depending on their spatial location. Figures \ref{red1805} and \ref{red1893} show the cells and the reddening values obtained by fitting the isochrones Girardi et al. 2002 in the respective cells,  using the same distance modulus and varying values of reddening. For IC~1805, $E(B-V)$ ranges from 0.7 -- 1.1 mag. For NGC~1893, the value of $E(B-V))$ ranges from 0.45 -- 0.65 mag. 
In the case of IC~1805, only a small region in the south-west region, shows a high value of extinction (1.1), the rest of the cluster shows 0.7 mag  a small region 0.8 mag. In the region of high extinction, there are a small number of stars which will not affect our analysis very strongly, as the mass functions are determined using mass bins of 0.5. Hence we use the mean value of 0.7 mag, as this will not change our results strongly.
For NGC~1893, the reddening varies from 0.45--0.65 mag , we have used a mean value of 0.5 for the determination of masses and mass functions as most of the stars lie in regions on $E(B-V)=0.5-0.55$ mag.

The observed data has been corrected for interstellar reddening using the coefficients given by  Dutra et al. (2002). 

\subsection{Radial Density Profiles}
\label{clusrad}
						     

\begin{figure} 
\centering
\includegraphics[width=8cm,height=7cm]{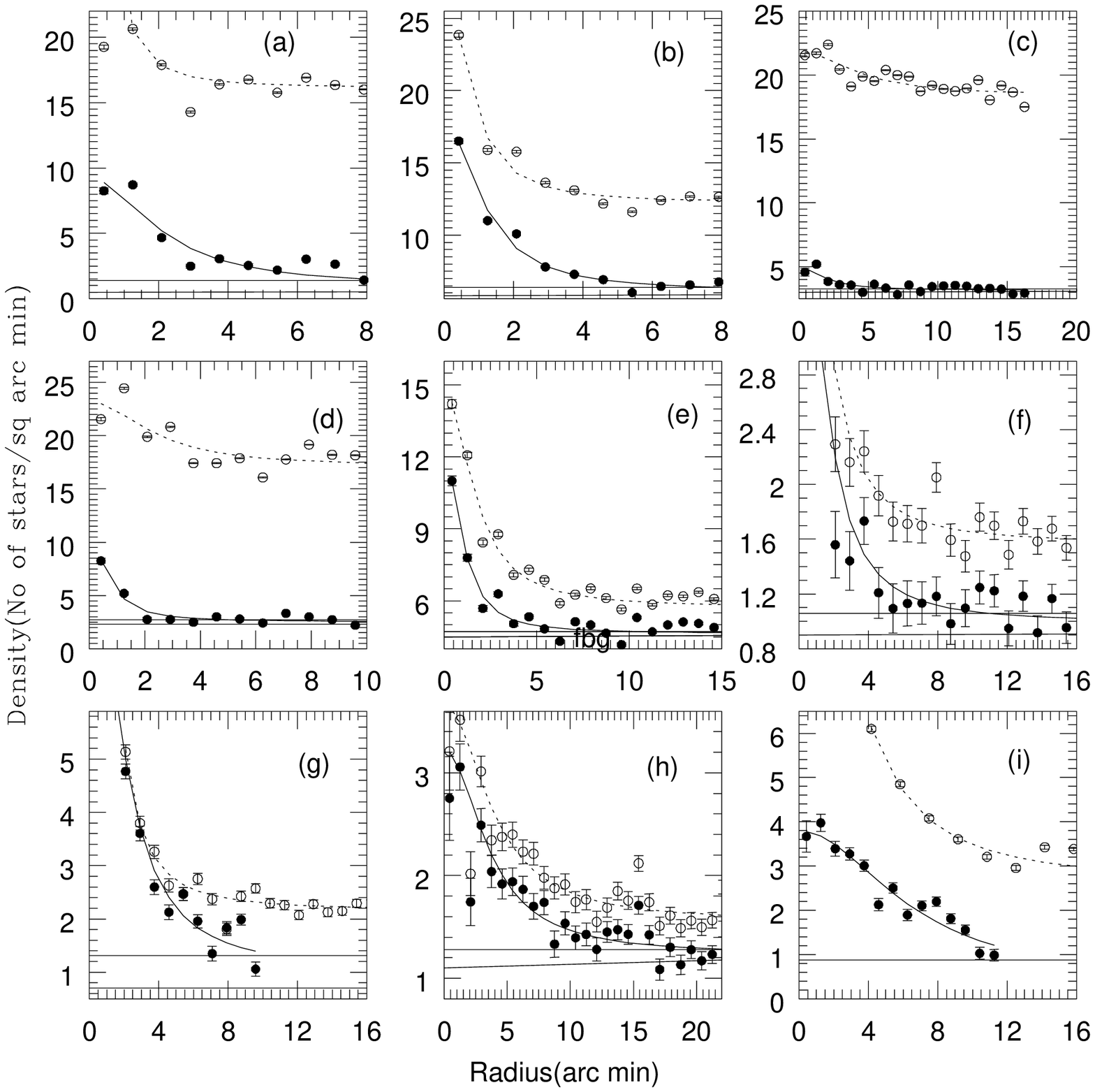}
\caption{Radial density profiles (a)NGC~6704 (b)NGC~6005 (c)NGC~6200 (d)NGC~6604 (e)IC~1805 (f)NGC~2286 (g)NGC~2489 (h)NGC~2354 and (i)NGC~1893}
\label{radall}
\end{figure}

For accurate determination of the cluster parameters, it is essential to determine the radial extent of clusters. As the 2MASS data offers all sky coverage we have the opportunity to study the outer regions of clusters.  The centres of the clusters are determined using a program described in Hasan et al. 2008.  A number of concentric circles with respect to the estimated centre are made in such a way that each annular region contains a significant number of stars. The number density of stars, $\rho_{i}$ in the $i^{th}$ region is calculated as $\rho_{i}=N_{i}/A_{i}$, where $N_{i}$ is the number of stars in the $i^{th}$ region of area $A_{i}$. 

Using the parameters obtained for the clusters, we use the method of Walker 1965  to find photometric members.  We then plot radial density profiles for possible photometric members as well as all the stars to  get the extent of the cluster. This is often very helpful especially in the case of the clusters which lie within the solar orbit and have very high field star densities  and where the  cluster stars are deeply embedded in the field.  The RDPs for the clusters using all stars (dotted line) and only those which satisfy the photometric criterion (solid line) are shown in the Fig.~\ref{radall}.  As is noticeable from the plots, a few of the clusters like NGC~6704, NGC~6005, NGC~6200, NGC~6604 and NGC~1893 are very faint and are only noticeable with this method. 

The $\chi ^{2}$ minimisation technique was used to fit the RDPs to the function $$\rho(r)=\frac{\rho_{0}}{1+(r/r_{c})^2}$$ King (1962) to determine $r_{c}$ and $\rho_{0}$. The cluster's core radius $r_{c}$ is the radial distance at which the value of $\rho(r)$ becomes half of the central density, $\rho_{0}$. The limiting radius of the cluster is the distance from the centre at which the star density becomes approximately equal to the field star density. The  sky coordinates of the cluster centres for epoch 2000, core Rad(core)and limiting radii Rad(lim) and background  and core density $\rho(bg), \rho(c)$ obtained by fitting to  King's profile are given in Table \ref{raddata}.
\begin{table*}
\small
\caption{Structural parameters from RDPs \label{raddata}}
\begin{tabular}{llllllllll}
\hline
Cluster&  RA(2000) & Decl.(2000) & $\rho(bg)$ & $\rho(c)$ & Rad(core) & Rad(lim) &  Rad(core) & Rad(lim)& $R_{GC}$  \\
       & (h:m:s) & (d:m:s) & stars/sq arc min& stars/sq arc min&($'$) & ($'$) & (pc)&(pc)&(kpc)\\
           \hline
 NGC 6704 & 18 50 45 & -05 12 18 & 0.95$\pm$ 0.43 &8.26 $\pm$ 0.88& 2.15$\pm$0.44&$8'$&1.8&6.7&6.1\\
 NGC 6005  & 15 55 48 & -57 26 12 &6.13$\pm$0.25 & 11.42$\pm$0.71&1.22$\pm$0.14&$6'$&0.8 &2.8&7.2\\
 NGC 6200 & 16 44 07 & -47 27 48 &3.17$\pm$0.1&1.81$\pm$0.33&2.03$\pm$0.64&$7'$&1.2&4.2&6.6\\
 NGC 6604 & 18 18 03 & -12 14 30 &2.54 $\pm$0.18& 7.33$\pm$0.99&0.79$\pm$0.18&$4.5'$&0.4& 2.2&6.9\\
 IC 1805  & 02 32 42 & +61 27 00 &4.64$\pm$0.08 &7.29$\pm$0.57&1.09$\pm$0.13&$9'$& 0.4  & 3.9 &9.6\\
 NGC 2286 & 06 47 40 & -03 08 54 &0.99$\pm$0.09  &3.15$\pm$0.33&1.63$\pm$0.29&$11'$&1.2&8.4&10.7\\
 NGC 2489 & 07 56 15 & -30 03 48 & 2.42$\pm$0.31 &7.83$\pm$0.44 & 2.11$\pm$0.25 & $10'$ &0.5  & 4.2&9.2 \\
 NGC 2354  & 07 14 10 & -25 41 24 &1.23$\pm$0.05 &2.01$\pm$0.18&3.65$\pm$0.48&$20'$&1.2&6.7& 9.2\\
 NGC 1893  & 05 22 44 &  +33 24 42&0.33$\pm$0.56 & 3.47$\pm$0.49&6.55$\pm$1.51 &$12'$& 3.1&12.7&12.1\\ 
 \hline
\end{tabular}
\end{table*}

To determine the  membership we use two criteria: the radial extent and the photometric criterion described by Walker 1965.  
The Walker method is valid only for main sequence stars while other luminosity classes and  groups require different methods for member identification. Hence, in this work, the results apply to the main sequence population of clusters under study. 

\subsection{Colour--magnitude diagrams}
\begin{figure}
\centering
\includegraphics[width=8cm,height=7cm]{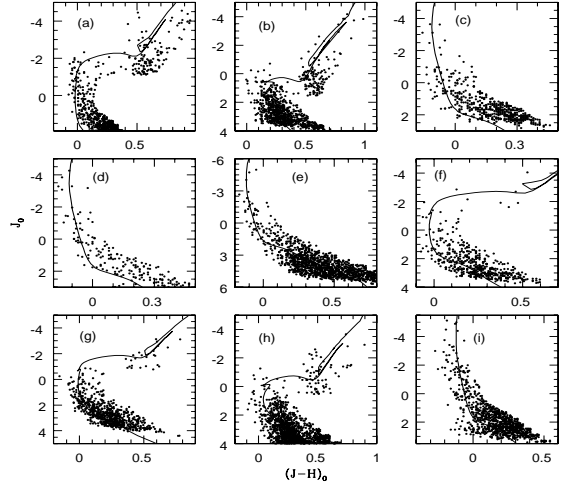}
\caption{Absolute CMDs (a)NGC~6704 (b)NGC~6005 (c)NGC~6200 (d)NGC~6604 (e)IC~1805 (f)NGC~2286 (g)NGC~2489 (h)NGC~2354 and (i)NGC~1893. Also plotted are the isochrones Girardi et al. 2002 for the clusters.}
\label{absall}
\end{figure}

The absolute CMDs for our cluster sample are shown in the Fig. \ref{absall}. 


\begin{figure}
\centering
\includegraphics[width=8cm,height=7cm]{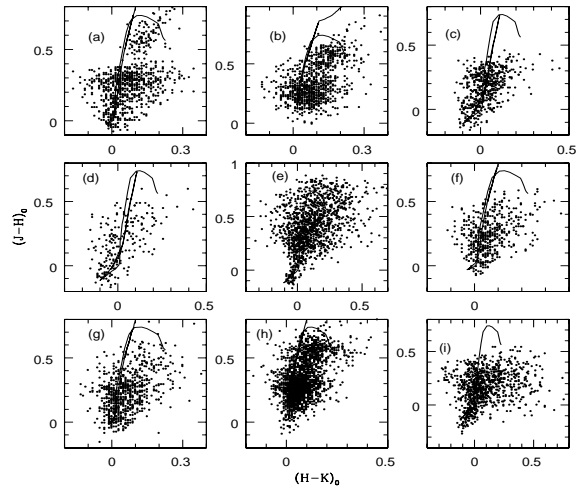}
\caption{Two--colour diagrams (a)NGC~6704 (b)NGC~6005 (c)NGC~6200 (d)NGC~6604 (e)IC~1805 (f)NGC~2286 (g)NGC~2489 (h)NGC~2354 and (i)NGC~1893}
\label{redall}
\end{figure}

The unreddened colour--colour diagrams $(J-H)_0$ versus $(H-K)_0$ for the photometric members of the  clusters are shown in the Fig.~\ref{redall}.

Table \ref{allpar} shows the values of the fundamental parameters of reddening, distance and age obtained for the clusters using isochrones Girardi et al. 2002 and compares them to those obtained by earlier authors. We have fit the isochrones to the `cleaned' CMD of the central regions of the clusers where field star contamination is minimised and then redone it for the entire extent of the cluster. In this work,  we are only referring to the population on the main sequence which does not have a very large age spread and therefore the use of single isochrone fit is justified. 

In the case of NGC~6704,  Forbes and DuPuy (1978) and Delgado et al. (1997) agreed on the distance, but disagreed on the age of the cluster basically due to the inclusion of giant stars as members. Delgado et al. (1997) included the giants and got a larger age of 200~Myr similar to the age of 250~Myr we obtained. In our case, for the cleaned CMD of the central region of the cluster, we got a large number of giant stars as probable members and inclusion of these led to the distance and age we obtained. These giant stars  appear very clearly in our `cleaned' CMD and lie in the central region of the cluster and hence are difficult to reject.  The distance estimate, however, agrees well with the value of 2974~pc in the Dias et al. 2007 catalogue.
In the case of NGC~6005,   Piatti et al. (1998) obtained similar reddening and ages to us, but differ strongly in the distance. Again in this case, this is because of the giant clump in the CMD, which we (and even the previous authors) have included as probable members. 
For NGC~6200,   Fitzgerald et al. (1977) obtained the distance based on photometry of 13 probable members and spectroscopy of 7 stars. Ours is based on a larger number of stars and hence can be considered an improvement on the previous value. This value, however agrees well with the value of 2054~pc in the Dias et al. 2007 catalogue.
The distances obtained for the clusters NGC 6604 by earlier authors and us perfectly agree. 
The distance estimates for IC~1805 are between 760~pc (Johnson 1968) to 2400 kpc (Sung and Lee 1995).  As we have used the method by  Walker 1965, we only identify main sequence members and our estimates are based on that population. Our values are within the range of estimates obtained by different authors.
The distance and age estimates obtained for NGC~2286 differ in this work and Pan et al. (1992).  The distance, however, agrees well with the value of 2600~pc in the Dias et al. 2007 catalogue.
In the case of NGC~2489, fitting the isochrones to the red giant members confirmed by     Piatti et al. (2007), we obtained a distance of 1445~pc and age 316~Myr compared to the values of 1800~pc and 500~Myr obtained by     Piatti et al. (2007).  
In the case of NGC~2354,  fitting the data obtained by us and the red giant members confirmed by Claria et al. (1999)
     , we obtained a difference in age  and distance estimates.
For NGC 1893, the distance obtained by Sharma et al (2007) 3250~pc, which is similar to the 3650~pc obtained by us.

\section{Luminosity and mass functions}

\begin{figure}
\includegraphics[width=10cm,height=10cm]{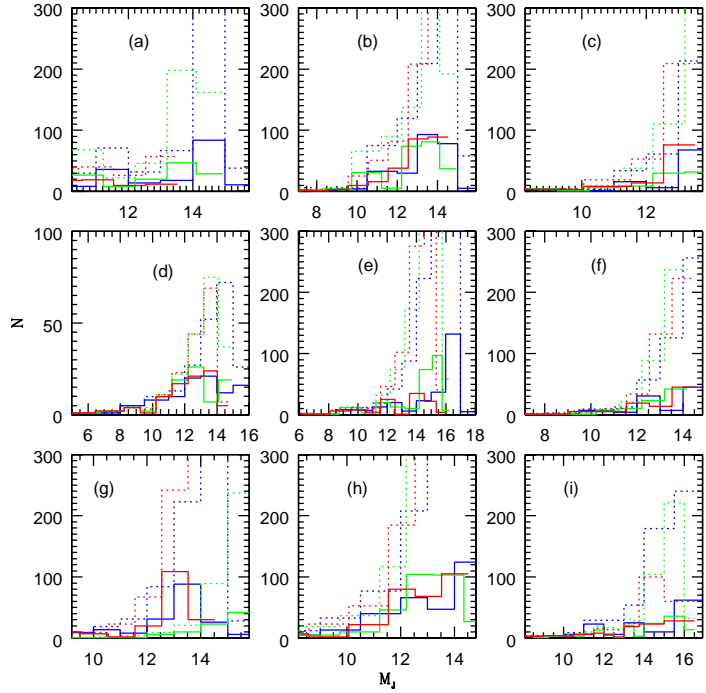}
\caption{Luminosity functions  (a)NGC~6704 (b)NGC~6005 (c)NGC~6200 (d)NGC~6604 (e)IC~1805 (f)NGC~2286 (g)NGC~2489 (h)NGC~2354 and (i)NGC~1893 ($J$ in blue, $H$ in green and $K$ in red, solid lines are the corrected luminosity functions and the dotted lines are the uncorrected luminosity functions)}
\label{lfall}
\end{figure}

The LFs obtained for clusters using observations have to be corrected for the following three factors: (i) fraction of cluster area studied (ii) completeness of data (iii) field star contamination. As the 2MASS data has 99.99\% completeness for the magnitude range used (see Table \ref{comp}) and we have  extracted data the complete cluster area, we only had to correct the LF for field star contamination. The LF was found for members based on the photometric criterion Walker 1965 in the $J$ vs $(J-H)$ plane using colour--magnitude filters. 
A similar colour--magnitude filter was applied for the apparent CMDs of the field area shown in Fig \ref{appall}. Thus, we obtain the approximate number of stars which are probable non-members, but still lie within our colour--magnitude filter. The number of field stars in each magnitude bin was then subtracted from the number of stars in the cluster area. The LFs in other bands were also found using  a similar method. Figure \ref{lfall} shows the uncorrected (dotted line) and corrected (solid line) LFs for the nine clusters in the $J$, $H$ and $K$ bands.

\begin{figure}
\includegraphics[width=8cm,height=6cm]{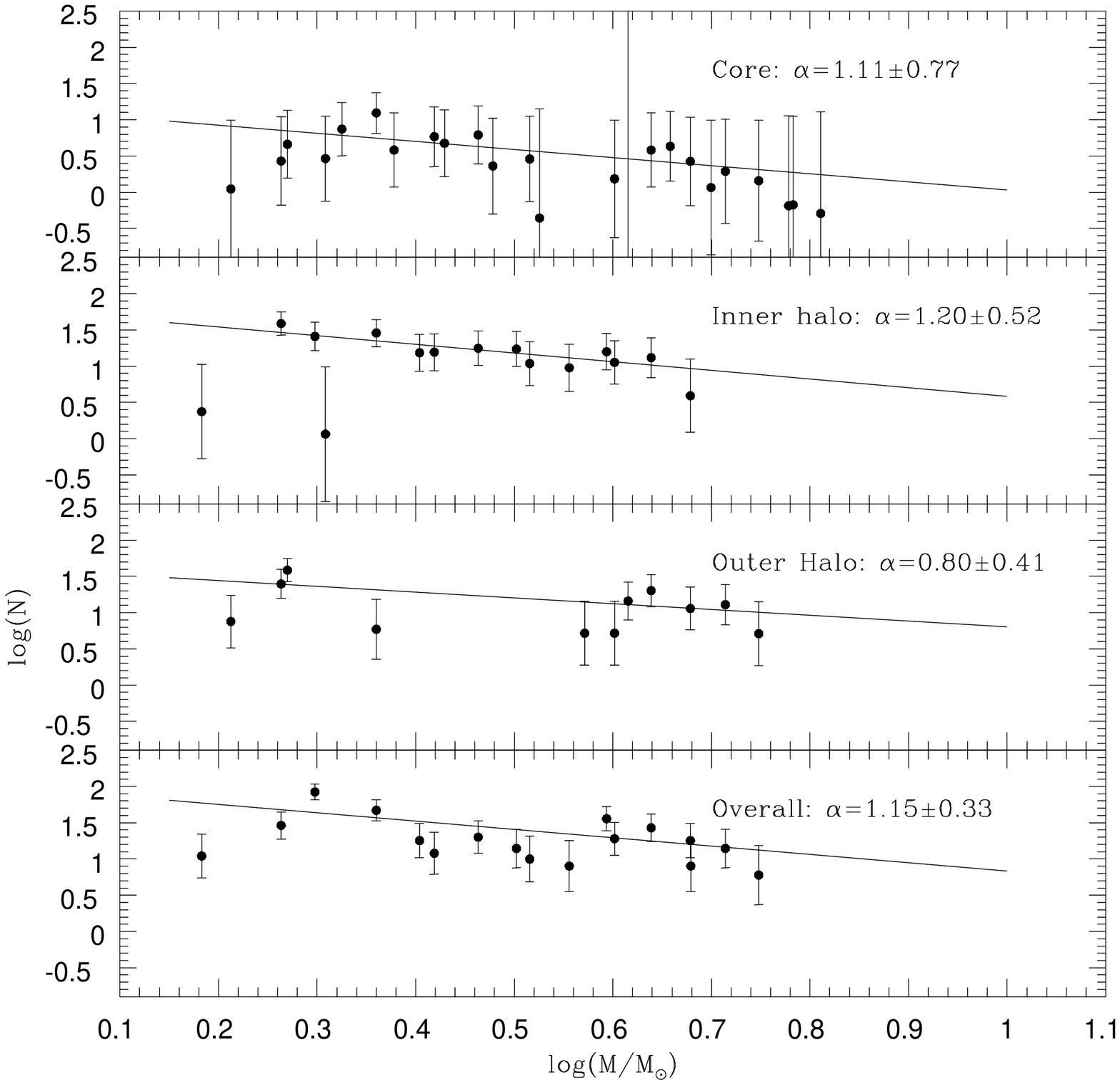}
\caption{NGC 6704: Mass function}
\label{mf6704}
\end{figure}

\begin{figure}
\includegraphics[width=8cm,height=6cm]{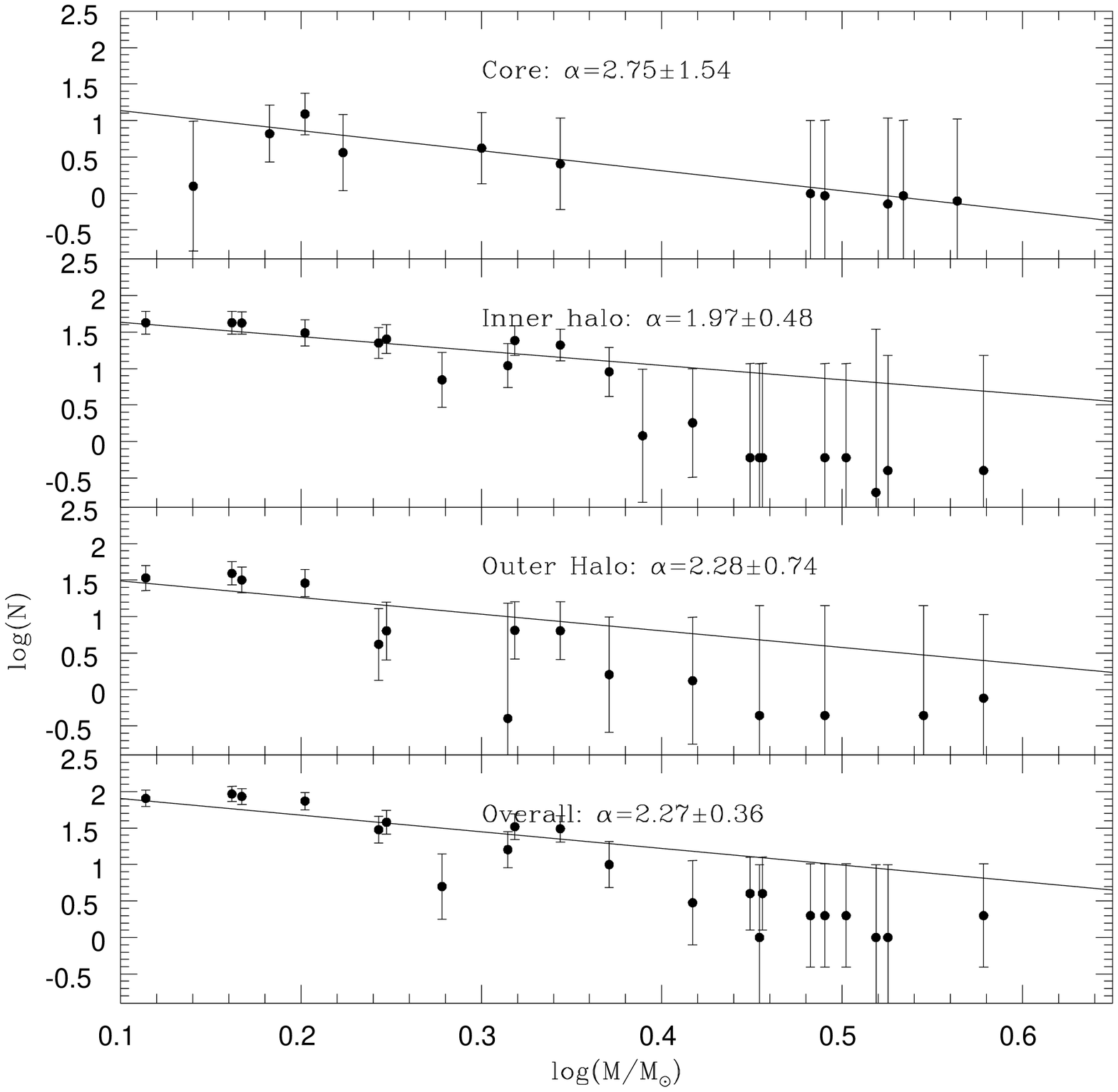}
\caption{NGC 6005: Mass function}
\label{mf6005}
\end{figure}

\begin{figure}
\includegraphics[width=8cm,height=6cm]{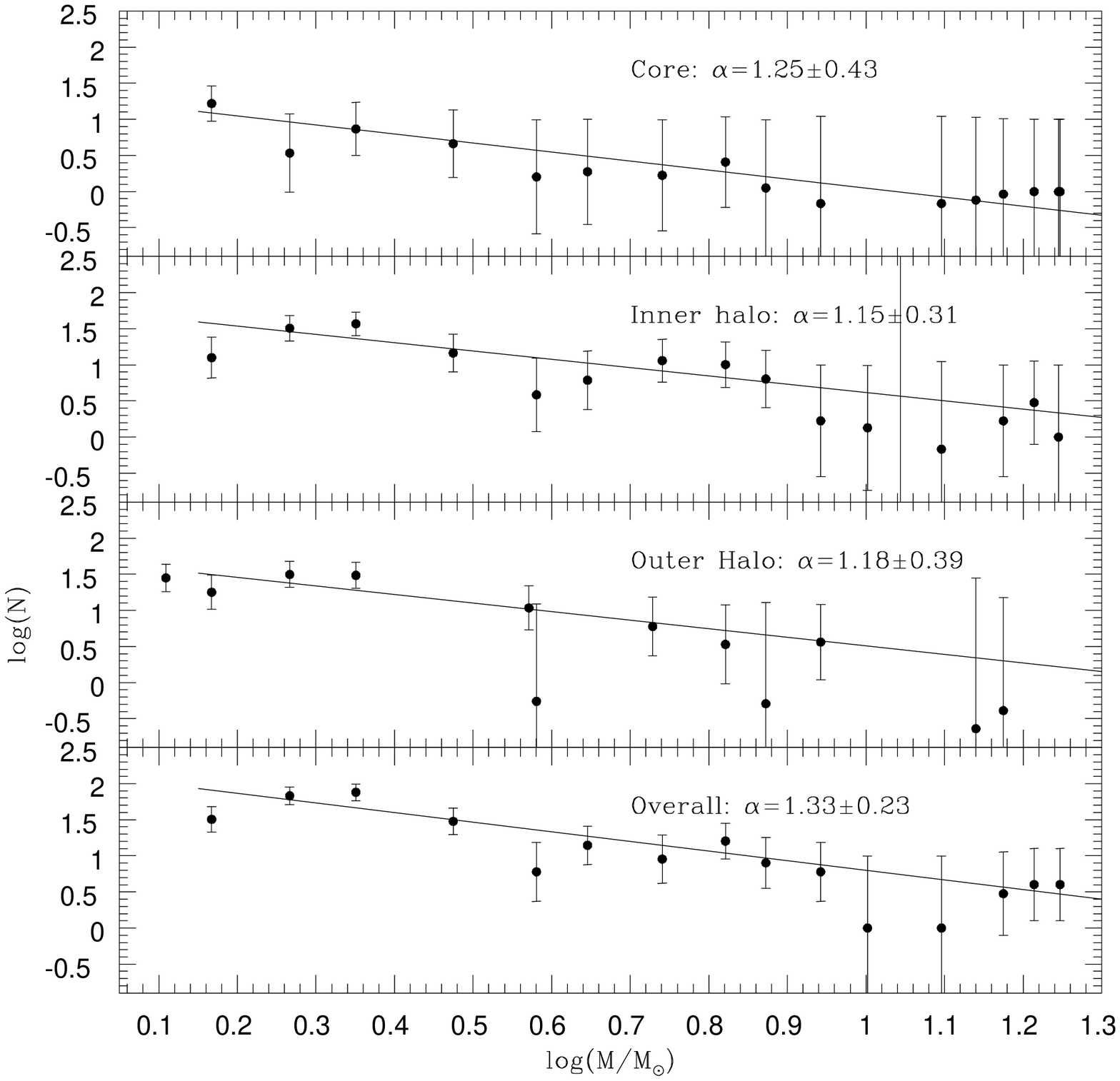}
\caption{NGC 6200: Mass function }
\label{mf6200}
\end{figure}

\begin{figure}
\includegraphics[width=8cm,height=6cm]{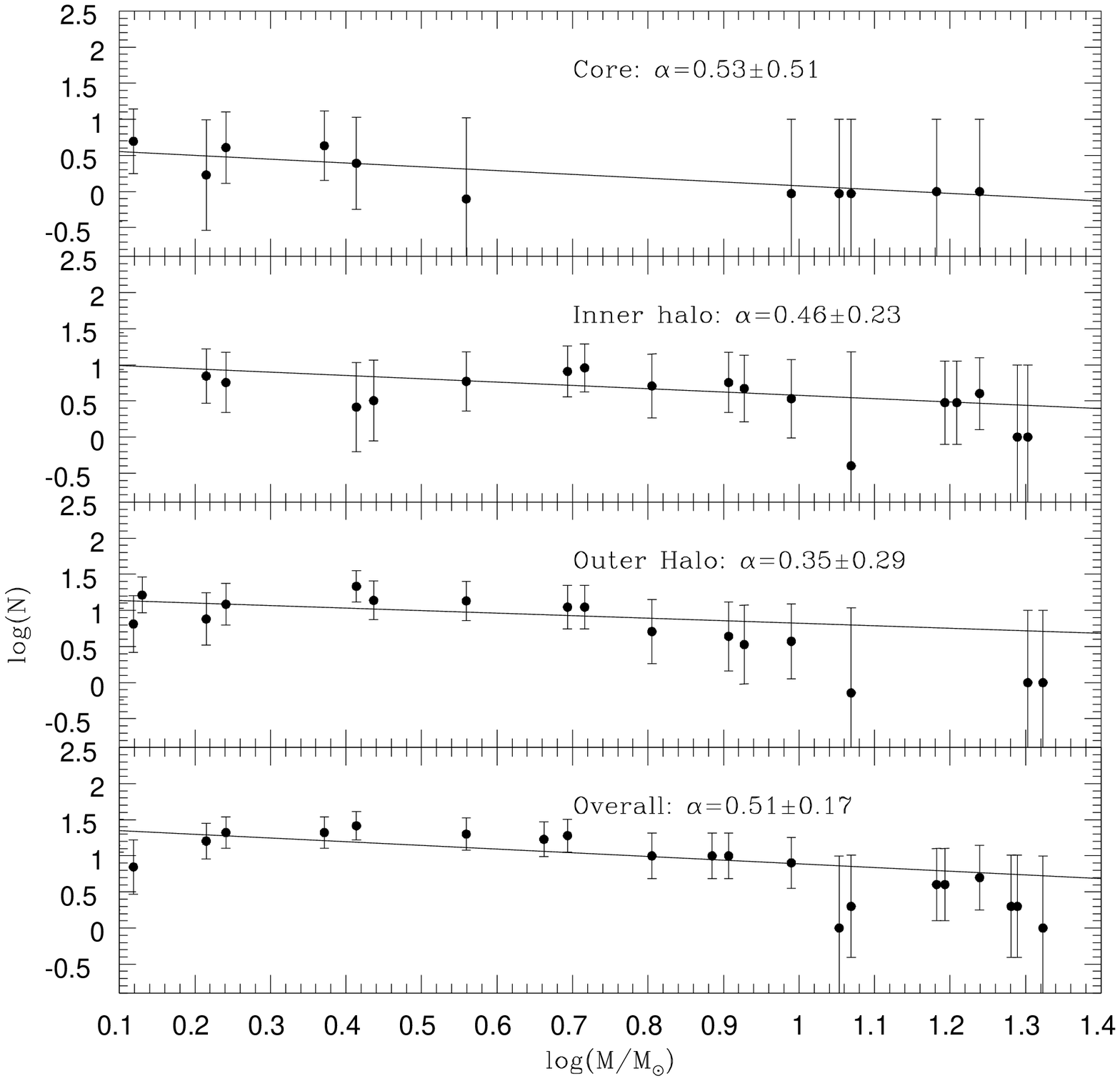}
\caption{NGC 6604: Mass function}
\label{mf6604}
\end{figure}

\begin{figure}
\includegraphics[width=8cm,height=6cm]{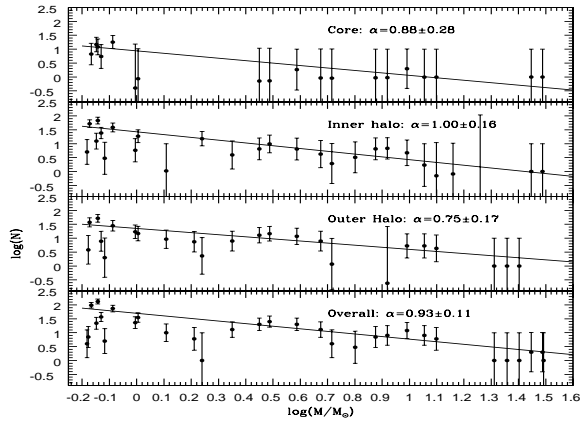}
\caption{IC 1805: Mass function}
\label{mf1805}
\end{figure}

\begin{figure}
\includegraphics[width=8cm,height=6cm]{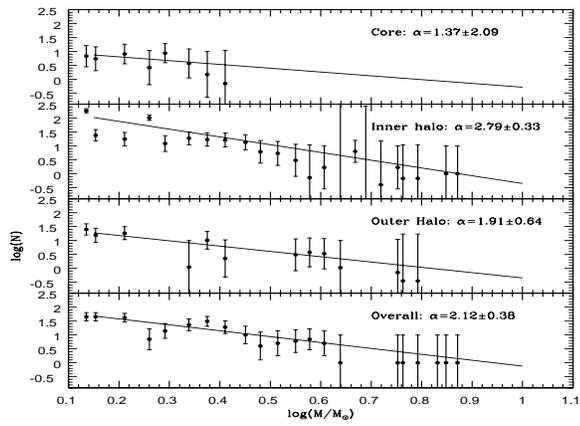}
\caption{NGC 2286: Mass function}
\label{mf2286}
\end{figure}

\begin{figure}
\includegraphics[width=8cm,height=6cm]{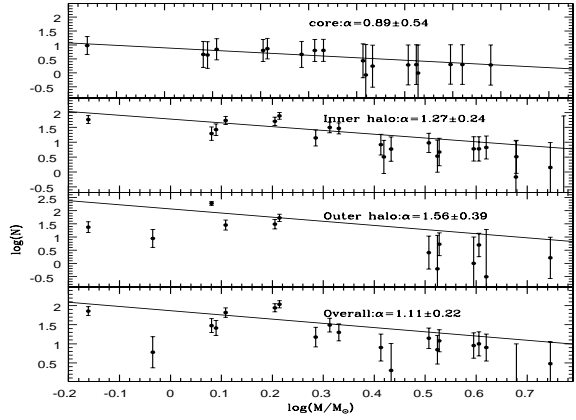}
\caption{NGC 2489: Mass function}
\label{mf2489}
\end{figure}

\begin{figure}
\includegraphics[width=8cm,height=6cm]{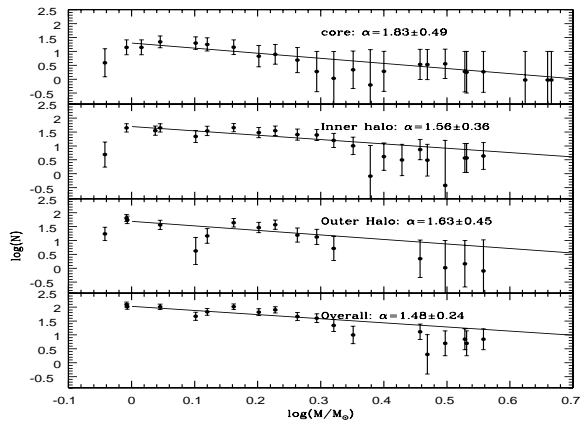}
\caption{NGC 2354: Mass function}
\label{mf2354}
\end{figure}

\begin{figure}
\includegraphics[width=8cm,height=6cm]{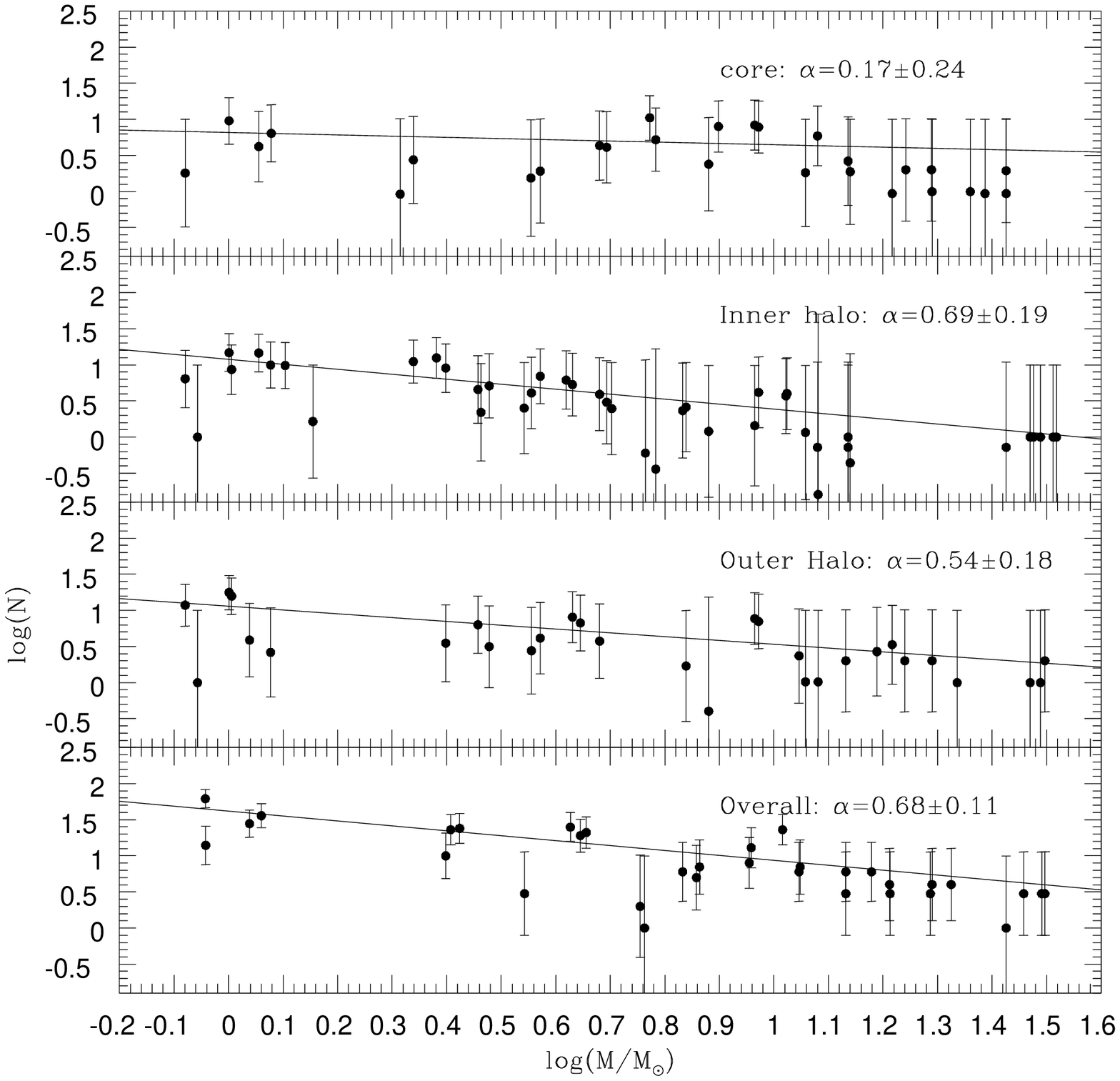}
\caption{NGC 1893: Mass function}
\label{mf1893}
\end{figure}

The MFs were constructed from the LFs using the isochrones Girardi et al. 2002 with the appropriate ages and distances and fitting them to a fourth order polynomial to find the mass--luminosity relation. The mass function, $\phi(M) = dN/dM \propto M^{-(\alpha)}$, is an indicator of the star formation process.  The relaxation times for the core and overall clusters have been calculated using the formula $t_{relax}=\frac{N}{8ln N}\times t_{cross}$ where $t_{cross}= R/\sigma_v$ is the crossing time, $N$ is the number of stars, $R$ is the radius and $\sigma_v$ is the velocity dispersion. We have used the value $\sigma_v$= 3 km s$^{-1}$ (Binney and Merriﬁeld 1998).. 

The clusters were divided into three regions (core, inner and outer halo) so as to obtain a significant number of stars in each region, shown in Table \ref{dypar}.

\begin{table*}
\small
\caption{Parameters estimated for NGC~6704, NGC~6005, NGC~6200, NGC~6604, IC~1805, NGC~2286, NGC~2489, NGC~2354 and NGC 1893}
\begin{tabular}{lllllll}
\hline
   Cluster & R (arc min) & $\Delta$m($M_{\odot}$) &$\alpha$  &  N  &  mass($M_{\odot}$ ) & $t_{relax}$(Myr) \\
             \hline
  NGC 6704  & &&    &       & \\ \hline
   core     & 0--2.15 &1.6--6 &1.11$\pm$0.77   & 59$\pm$31     & 19$\pm$10&\\
    halo1  &  2.15--5&1.5--9.4  &1.20$\pm$0.52    &212$\pm$140       &330$\pm$218&\\
   halo2  & 5--8 &1.6--11.7&  0.80$\pm$0.41 & 325$\pm$267     &  110$\pm$90 &\\
  overall       & 0--8&1.5--11.7 &  1.15$\pm$0.33     &  596$\pm$437    &  260$\pm$190&26\\ \hline
 
 NGC 6005  & &&   &       & \\ \hline
   core     & 0--1.22 &1.4-3.7&2.75$\pm$1.54 &33$\pm$11   & 12$\pm$4 & \\
    halo1  &  1.22--4& 1--3.8& 1.97$\pm$0.48    &   402$\pm$252    &  348$\pm$218 &\\
   halo2  & 4--6 &1--3.8& 2.28$\pm$0.74  &   435$\pm$352    &  119$\pm$96& \\          
  overall       &0--6 & 1--3.8&  2.27$\pm$0.36     &  866$\pm$629    &  381$\pm$276 &15\\ \hline

NGC 6200  & &&    &       & \\ \hline
   core     & 0--2.03 &1.5--17.7& 1.25$\pm$0.43   & 57$\pm$32     & 64$\pm$36&0.7\\
    halo1  &  2.03--4.5&1.5--17.6&  1.15$\pm$0.31    & 175$\pm$131      &  288$\pm$215 &\\
   halo2  & 4.5--7 &1.3--15&  1.18$\pm$0.39  &  219$\pm$153    &  326$\pm$227 &\\          
  overall       &0--7& 1.5--17.7 &  1.33$\pm$0.23    &  479$\pm$397    &  503$\pm$417 &13.8\\ \hline

NGC 6604  & &&    &       & \\ \hline
   core     & 0--0.79 &1.2--17.3& 0.53$\pm$0.51   & 17$\pm$3     & 34$\pm$6&0.1\\
    halo1  &  0.79--2.6&1.2--27&  0.46$\pm$0.23    &  50$\pm$13     &  131$\pm$347 \\
   halo2  & 2.6--4.5 & 1--21 &0.35$\pm$0.29  &   118$\pm$70   &  202$\pm$119 &\\          
  overall       & 0--4.5 &1.2--19.5&  0.51$\pm$0.17     &   200$\pm$110   &  304$\pm$167&3.58 \\ \hline
 
 IC 1805 & &&  &       & \\ \hline
   core     & 0--1.08 &0.7--31& 0.88$\pm$0.28   &  39$\pm$6    & 81$\pm$12&0.2\\
    halo1  &  1.08--5&0.7--31&  1.00$\pm$0.16    &  413$\pm$301     &  324$\pm$236 &\\
   halo2  & 5--9 &0.7--25&  0.75$\pm$0.17  &   799$\pm$702   &  196$\pm$172& \\          
  overall       &0--9& 0.7--31 &  0.93$\pm$0.11     &   1256$\pm$1017   &  414$\pm$335 &29\\ \hline

NGC 2286 & &&    &       & \\ \hline
   core     & 0--1.63 &1.4--2.6& 1.37$\pm$2.09   &  17$\pm$5    & 6$\pm$2&0.32\\
    halo1  &  1.63--6.5&1--7.4&  2.79$\pm$0.33    &  146$\pm$91     &  348$\pm$217& \\
   halo2  & 6.5--11 &1.2--6.2&  1.91$\pm$0.64  &  262$\pm$235    &  38$\pm$34& \\          
  overall       & 0--11 &1.4--7.4&  2.12$\pm$0.38  &   363$\pm$279      &  150$\pm$115&18 \\ \hline

NGC 2489  & &&    &       & \\ \hline
   core     & 0--1.1 &0.7--4.8& 0.89$\pm$0.54   & 30 $\pm$5    & 26$\pm$4&0.18\\
    halo1  &  1.1--5.5&0.7--5.9&  1.27$\pm$0.24    &   264$\pm$123      &  442$\pm$206 &\\
   halo2  & 5.5--10 &0.7--5.9&  1.56$\pm$0.39  & 341$\pm$293     &  212$\pm$182 &\\          
  overall       & 0--10 &0.7--5.6&  1.11$\pm$0.22    &   709$\pm$532   &  264$\pm$198& 19.4\\ \hline
 
 NGC 2354 & &&    &       & \\ \hline
   core     & 0--3.65 &1--4.6& 1.83$\pm$0.49   & 98$\pm$46     & 34$\pm$16&\\
    halo1  &  3.65--12&1--3.6&  1.56$\pm$0.36    &  641$\pm$504     &  96$\pm$75& \\
   halo2  & 12--20 &1--3.6&  1.63$\pm$0.45  &  1077$\pm$976    &  80$\pm$72 &\\          
  overall       &0--20& 1--3.6 &  1.48$\pm$0.24     &  1834$\pm$1541    &  215$\pm$181 &70\\ \hline

NGC 1893  & &&    &       & \\ \hline
   core     & 0--3 &0.8--27& 0.17$\pm$0.24   &  80$\pm$46    & 89$\pm$51&2.4\\
    halo1  &  3--7&0.8--33&  0.69$\pm$0.19    &   258$\pm$218    &  119$\pm$100 &\\
   halo2  & 7--12 &0.8--33&  0.54$\pm$0.18  &  525$\pm$ 503   &  268$\pm$256 &\\          
  overall       &0--12& 0.8--31 &  0.68$\pm$0.11   &  827$\pm$744    &  365$\pm$328& 67\\ \hline
\end{tabular}
\label{dypar}
\end{table*}


  Table \ref{dypar} also shows the  values of the mass estimates and $\alpha$ for different regions of the clusters which are indicative of mass segregation. The mass estimates for the clusters  are the lower limits of the masses for these clusters, as a large fraction of the mass lies in low mass stars which are embedded in the field. The number of stars $N$ in the table are given with the errors which are equal to the number of stars present in the proportionate region of the field.

Figure \ref{mf6704} shows the mass function for the cluster NGC~6704 where the $\alpha$ value  was found to be 1.15$\pm$0.33  for the overall cluster, 1.11$\pm$0.77 in the core region, 1.20$\pm$0.52  in halo1 and 0.80$\pm$0.41 in halo2.
 The relaxation time is 26~Myr for the overall cluster. The age of the cluster based on the isochrone fit is 250~Myr and the age based on the most massive star on the main sequence (3.9 $M_{\odot}$) is $\leq$ 330~Myr. Hence the cluster has dynamically relaxed ($\tau \approx 9$). Some of the less massive stars have moved to the outer regions of the cluster and have been lost from halo2 and hence halo2 has a flatter value of $\alpha$. Halo1 has a larger number of low mass stars which will slowly be lost as they move to the halo2. 

 In the case of NGC~6005 (Fig. \ref{mf6005}), the $\alpha$ value of the MF has been found in the core, halo1 and halo2 as 2.75$\pm$1.54, 1.97$\pm$0.48 and 2.28$\pm$0.74 respectively. The cluster has an age of 1258~Myr and has an overall $\alpha$ value of $2.27\pm0.36$. The relaxation time for NGC~6005 is 15~Myr and $\tau \approx 83$. Significant mass segregation must have already taken place in the cluster but many of the massive stars of this cluster have already moved away from the main sequence (as seen in the CMD). These stars have lost mass and moved to the outer regions and the many low mass stars have been lost due to evaporation in the presence of strong Galactic tidal forces. This is also evident from the small size of the cluster (2.8~kpc). The most massive star on the main sequence has a mass of $2 M_{\odot}$, with a nuclear age of 1800~Myr. 
This cluster is a good example of a segregated cluster with high values of $\alpha$ which shows the effect of both aspects: dynamics and evolution of stars.

For the cluster NGC~6200 (Fig. \ref{mf6200}), the relaxation times for the core and overall cluster are 0.7~Myr and 13.8~Myr respectively. The $\alpha$ value of the core 1.25$\pm$0.43 shows that the core has relaxed since the cluster has an age of 6.3~Myr. However, halo1 and halo2 are in the process of relaxation and hence their $\alpha$ values are  1.15$\pm$0.31 and  1.18$\pm$0.39 respectively. The overall cluster has $\alpha=$ 1.33$\pm$0.23 as the cluster has partially relaxed.

NGC~6604 has an age of 6.3~Myr which exceeds the relaxation times for the core (0.1~Myr) and cluster (3.58~Myr) respectively. Hence, the cluster has relaxed and has $\alpha$ values 0.53$\pm$0.51,  0.46$\pm$0.23,  0.35$\pm$0.29 and  0.51$\pm$0.17 for the core, halo1, halo2 and overall cluster respectively.  Since the age of the cluster exceeds the relaxation time, significant relaxation/mass segregation would have taken place as is evident from the similar values of alpha for the core and the inner and outer halos.
 
 IC~1805 has an age of 4~Myr and the relaxation times for the core and overall cluster are 0.2~Myr and 29~Myr respectively. If we assume a Salpeter IMF ($\alpha$=2.35), we see that the mass function of the cluster, seems to have changed as is evident from the $\alpha$ values of the core (0.59$\pm$0.17), halo1 (0.88$\pm$0.14), halo2 (0.68$\pm$0.02) and  overall cluster (0.69$\pm$0.14). This indicates an excess of high mass stars in the overall cluster and also in the core compared to the inner halo, indicative of a high degree of mass segregation. This has been earlier reported by Sagar et al 1988. 

NGC~2286 has an age of 200~Myr which exceeds the core and overall relaxation times of 0.32~Myr and 18~Myr. The $\alpha$ values of the core, halo1, halo2 and overall cluster are 1.37$\pm$2.09, 2.79$\pm$0.33, 1.91$\pm$0.64 and 2.12$\pm$0.38  respectively, showing that the mass seggregation process must have taken place, but many of the high mass stars have moved away from the main sequence and have lost mass.

NGC~2489 is an old relaxed cluster of age 316~Myr which is much larger compared to its relaxation time of the core of 0.18~Myr and overall cluster 19.4~Myr. This is evident from the flat $\alpha$ value of the core (0.89$\pm$0.54). The $\alpha$ values of halo1, halo2 and overall cluster are 1.27$\pm$0.24, 1.56$\pm$0.39 and 1.11$\pm$0.22 respectively.

 NGC~2354 has an age of 630~Myr which is large compared to the overall relaxation time of 70~Myr. The cluster core has a $\alpha$ value of 1.83$\pm$0.49. The halos and overall cluster have similar $\alpha$ values of 1.56$\pm$0.36, 1.63$\pm$0.45  and 1.48$\pm$0.24  respectively. As seen in the CMD, the cluster is old and most massive stars have evolved away from the main sequence and hence the core has a larger number of low mass stars.

NGC~1893 is a very young cluster of age 4~Myr which shows signs of overall mass segregation not only in the core which has a relaxation time of 2.4~Myr, but also in the overall cluster whose relaxation time is very large (67~Myr). The $\alpha$ values for the core, halo1, halo2 and overall cluster are 0.17$\pm$0.24, 0.69$\pm$0.19, 0.54$\pm$0.18  and  0.68$\pm$0.11 respectively. This cluster also shows signs of early  mass segregation as the relaxation time of the cluster clearly exceeds the age of the cluster. Sharma et al (2007) also obtained results suggesting primordial mass segregation in this cluster. This cluster is  located in the Galactic anticenter region at a distance of $\approx$ 14.5~kpc from the Galactic centre. Using Spitzer observations, \cite{cara08} found the maximum mass of stars in the cluster to be $28-46 M_{\odot}$ and infer that the cluster does not show any peculiarity regarding the ongoing star formation. 

 \section{Conclusions}
 In this paper, using 2MASS data, we have studied mass segregation in nine clusters in diverse environments to understand their structure and dynamics. The RDPs of the clusters have been plotted  (Fig. \ref{radall}) and the parameters for the clusters such as reddening, distance and age have been determined using isochrone fits (Table \ref{allpar}). We have also plotted the LFs in the $J$, $H$ and $K$ bands and used the derived mass--luminosity relation to find the MFs using all three bands independently (see Figs \ref{lfall}--\ref{mf1893}). Clusters have been divided into three regions: core, inner and outer halo. The $\alpha$ values have been determined for different regions and the overall clusters as a function of the parameter $\tau$.  We use the change in $\alpha$ values for different regions to estimate the level of mass segregation of the clusters. 

\begin{figure}
\includegraphics[width=8cm,height=6cm]{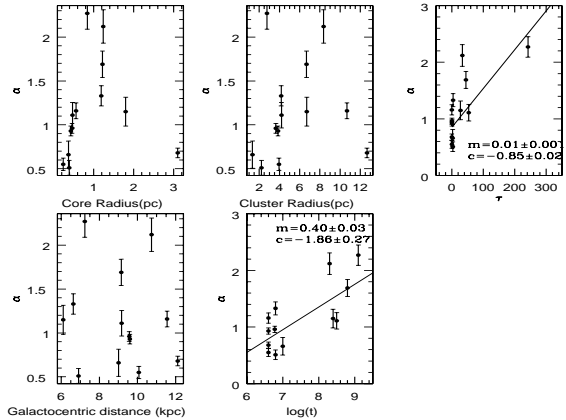}
\caption{Dependence of $\alpha$ on various parameters}
\label{chi}
\end{figure}

The $\alpha$ values of mass functions of the clusters under study range from 0.17 to 2.79. Figure \ref{chi} shows the dependence of $\alpha$  of clusters as a function of various parameters for 13 clusters (9 from this work and 4 from Hasan et al. 2008). Though our sample is small, it is homogeneous, in the sense of photometric data as well as methods of data analysis thus making it a controlled sample. Such studies are not suitable using heterogeneous datasets where unknown biases may be present. 

It is interesting to note a very high confidence level in the correlation of $\alpha$ with age and $\tau$. As clusters age, they have steeper values of $\alpha$. The value of $\alpha$ value increases with age and $\tau$ and fits straight lines with slopes $m$ and y-intercepts $c$ given by $m=0.40\pm0.03$, $c=-1.86\pm0.27$ and $m=0.01\pm0.001$, $c=-0.85\pm0.02$, respectively. The increase in the value of $\alpha$  with age and $\tau$,  is a clear indicator of the dynamical processes involved where mass segregation can be explained by dynamics. The confidence level of the  Pearson's product-moment correlation of $\alpha$  with age is 0.76 with p=0.002 and with $\tau$ is 0.71 with p=0.007. \footnote{The p value shows at what level of confidence the null hypothesis (correlation) can be rejected. For example, p=0.05 shows a 95\% probability that the hypothesis of a correlation is correct.}  The value of $\alpha$  increases with Galactocentric distance, indicating a larger number of low mass stars in clusters at larger Galactocentric distances due to lesser evaporation of stars.

The cluster NGC~6704 had an $\alpha$ value of 1.15$\pm$0.33 for the overall cluster with an age exceeding 9 times the relaxation time. The cluster has dynamically relaxed, many of the less massive stars have moved to the outer regions of the cluster, some have been lost due to evaporation and hence halo2 has a flatter value of $\alpha$ compared to halo1. 
NGC~6005 is an old cluster which has been mass segregated and has high values of $\alpha$ due to the effect of both dynamics and evolution of stars, in which massive stars have evolved, lost mass and moved to the outer regions of the cluster. 
In the case of the cluster NGC~6200, the relaxation times for the core and cluster as a whole are 0.7~Myr and 13.8~Myr respectively and the cluster  has partially relaxed. The $\alpha$ value of the core is 1.25$\pm$0.43 and it shows that the core has a larger number of high mass stars due to relaxation since the cluster has an age of 6.3~Myr ($> t_{relax}$ for the core). However, the inner and outer halos are in the process of relaxation and their $\alpha$ values are  1.15$\pm$0.31 and  1.18$\pm$0.39 respectively. 

NGC~6604, though young, has an age of 6.3~Myr which exceeds the relaxation times for the core (0.1~Myr) and cluster (3.58~Myr) respectively. Hence, the cluster has relaxed and has $\alpha$ values 0.53$\pm$0.51, 0.46$\pm$0.23,  0.35$\pm$0.29 and  0.51$\pm$0.17  for the core, halo1, halo2 and overall cluster respectively. 
 
 IC~1805 has an age of 4~Myr and the relaxation times for the core and overall cluster are 0.2~Myr and 29~Myr respectively. It already shows mass segregation as earlier reported by Sagar et al 1988.   The $\alpha$ values of  the mass function of the cluster, are core (0.59$\pm$0.17), halo1 (0.88$\pm$0.14), halo2 (0.68$\pm$0.02) and  overall cluster (0.69$\pm$0.14). 
NGC~2286 has an age of 200~Myr which exceeds the core and overall relaxation times of 0.32~Myr and 18~Myr. The $\alpha$ values of the core, halo1, halo2 and overall cluster are 1.37$\pm$2.09, 2.79$\pm$0.33, 1.91$\pm$0.64 and 2.12$\pm$0.38 respectively. Mass seggregation process must have taken place, but many of the high mass stars have moved away from the main sequence and have lost mass and the outer halo seems to have lost low mass stars  and hence has a flatter $\alpha$ .
. 

NGC~2489 is an old relaxed cluster  and many of the low mass stars from the core have moved to the outer regions of the cluster. This is evident from the flat $\alpha$ value of the core (0.89$\pm$0.54) and the larger  $\alpha$ values of halo1, halo2 and overall cluster (1.27$\pm$0.24, 1.56$\pm$0.39 and 1.11$\pm$0.22 respectively).
 
 NGC~2354 is an old cluster and most massive stars have evolved away from the main sequence and the halos and overall cluster have similar $\alpha$ values of 1.56$\pm$0.36, 1.63$\pm$0.45  and 1.48$\pm$0.24   respectively. 

NGC~1893 is a very young cluster of age 4~Myr which shows signs of overall mass segregation not only in the core which has a relaxation time of 2.4~Myr, but also in the overall cluster whose relaxation time is very large (67~Myr). The $\alpha$ values for the core, halo1, halo2 and overall cluster are 0.17$\pm$0.24, 0.69$\pm$0.19, 0.54$\pm$0.18  and  0.68$\pm$0.11 respectively.

Of the nine clusters studied, two clusters (IC~1805 and NGC~1893), are too young to be dynamically relaxed and we speculate this as evidence for primordial mass segregation.  Mass segregation by birth is a natural expectation because protostars near the density centre of the cluster have more material to accrete. The actual efficiency of this mechanism is still a matter of debate is still a matter of debate (Krumholz et al. 2005; Krumholz and Bonnell 2009). McMillan et al. (2007) presented an alternative scenario for a dynamical origin of early mass segregation in young clusters. Even if the clumps are not initially segregated, if their internal segregation timescale is shorter than the time needed for the clumps to merge, they will segregate through standard two-body relaxation and preserve this segregation after they have merged. The multiscale dynamical evolution of clumpy systems is, in this case, responsible for rapidly leading to mass segregation in young clusters without invoking any mechanism associated with the star-formation process. Recent simulations by the star-formation process. Recent simulations by Allison
et al. (2009, 2010) showed that early mass segregation can
be due to dynamical effects even in timescales as short as a
Myr, thus not requiring the need of primordial mass segre-
gation which would violate the universality of the IMF and
set constraints on the origin of the IMF.
     Understanding the origin of mass segregation can also
help diﬀerentiate between possible models of massive star
formation. Do massive stars form in the centres of clusters,
or do they migrate there over time due to gravitational in-
teractions with other cluster members? In particular, are
the masses of the most massive stars set by the mass of the
core from which they form (Krumholz and Bonnell 2009) or
by competitively accreting mass due to being located at a
favourable position in the cluster (Bonnell and Davies 1998;
Krumholz et al. 2005; Bonnell and Bate 2006)? Allison et al.
(2009) showed that dynamical mass segregation can occur on
a few crossing timescales suggests that massive stars could
form in relative isolation in large cores and mass segregate
later, possibly avoiding the need for competitive accretion
as dominant process to form the most massive stars in the
centre of a cluster. However, the simulations by Moeckel
and Bonnell (2009) show that for such young systems, star
formation scenarios predicting general primordial mass seg-
regation are inconsistent with observed segregation levels.
They found that a star-formation scenario in which only the
most massive stars are primordially segregated is consistent
with observations, and oﬀers a way to account for compact
groups of young, massive stars.
     Currently we cannot say conclusively if mass segrega-
tion is a birth phenomenon (Gouliermis et al. (2004), or
whether the more massive stars form anywhere throughout
the proto-cluster volume. Star clusters that have already
blown out their gas at ages of one to a few Myr are typi-
cally mass- segregated (e.g. R136, Orion Nebula Cluster).
Assuming primordial mass segregation would imply that
massive stars ( > 10M⊙ ) only form in rich clusters, and
reject the possibility they can also form in isolation (see Li
et al. (2003); Parker and Goodwin (2007)).  A better understanding of the effects of dynamical evolution is required to clearly differentiate between present dynamically derived star cluster properties and those which were imprinted by star-formation processes.

\section{Acknowledgements}
%

We would like to thank the referee for his valuable suggestions. This publication makes use of data products from the Two Micron All Sky Survey, which is a joint project of the University of Massachusetts and the Infrared Processing and Analysis Centre/California Institute of Technology, funded by the National Aeronautics and Space Administration and the National Science Foundation. This research has made use of the WEBDA database, operated at the Institute for Astronomy at the University of Vienna (http://www.univie.ac.at/webda) founded by J.-C.Mermilliod (1988, 1992)  devoted to observational data on Galactic open clusters.) Virtual observatory tools like Aladin and Topcat have been used in the analysis.
This research has been funded by the Department of Science and Technology (DST), India under the Women Scientist Scheme (PH).



\label{lastpage}

\end{document}